\documentclass[journal=jacsat,manuscript=article, BackFigs]{achemso}

\usepackage{chemformula} 
\usepackage{graphicx}
\usepackage{epstopdf}
\usepackage{bm,amsmath,amssymb} 
\usepackage{color, soul} 
\usepackage{dcolumn}   
\usepackage{multirow}

\newcommand{\etal}{{\em et al.\ }}

\author{Jiawei Huang}
\affiliation{Key Laboratory for Quantum Materials of Zhejiang Province, Department of Physics, School of Science, Westlake University, Hangzhou, Zhejiang 310030, China}
\author{Changming Ke}
\affiliation{Key Laboratory for Quantum Materials of Zhejiang Province, Department of Physics, School of Science, Westlake University, Hangzhou, Zhejiang 310030, China}
\alsoaffiliation{Institute of Natural Sciences, Westlake Institute for Advanced Study, Hangzhou, Zhejiang 310024, China}
\author{Wei Zhu}
\affiliation{Key Laboratory for Quantum Materials of Zhejiang Province, Department of Physics, School of Science, Westlake University, Hangzhou, Zhejiang 310030, China}
\alsoaffiliation{Institute of Natural Sciences, Westlake Institute for Advanced Study, Hangzhou, Zhejiang 310024, China}
\author{Shi Liu}
\email{liushi@westlake.edu.cn}
\affiliation{Key Laboratory for Quantum Materials of Zhejiang Province, Department of Physics, School of Science, Westlake University, Hangzhou, Zhejiang 310030, China}
\alsoaffiliation{Institute of Natural Sciences, Westlake Institute for Advanced Study, Hangzhou, Zhejiang 310024, China}

\title{One Dimensional Ferroelectric Nanothreads with Axial and Radial Polarization}%

\begin{document}

\newpage
\begin{abstract}
Long-range ferroelectric crystalline order usually fades away as the spatial dimension decreases, hence there are few two-dimensional (2D) ferroelectrics and far fewer one-dimensional (1D) ferroelectrics. Due to the depolarization field, low-dimensional ferroelectrics rarely possess the polarization along the direction of reduced dimensionality. Here, using first-principles density functional theory, we explore the structural evolution of nanoribbons of varying widths constructed by cutting the 2D sheet of ferroelectric $\alpha$-III$_2$IV$_3$ (III=Al, Ga, In; VI=S, Se, Te).
We discover a one-dimensional ferroelectric nanothread (1DFENT) of ultrasmall diameter with both axial and radial polarization, potentially enabling ultra-dense data storage with a 1D domain of just three unit cells being the functional unit. The polarization in 1DFENT of Ga$_2$Se$_3$ exhibits an unusual piezoelectric response: a stretching stress along the axial direction will increase both axial and radial polarization, referred to as auxetic piezoelectric effect. Utilizing the intrinsically flat electronic bands, we demonstrate the coexistence of ferroelectricity and ferromagnetism in 1DFENT and a counterintuitive charge-doping-induced metal-to-insulator transition. The 1DFENT with both axial and radial polarization offers a counterexample to the Mermin--Wagner theorem in 1D and suggests a new platform for the design of ultrahigh-density memory and the exploration of exotic states of matter.
\end{abstract}
\newpage
\section{Introduction}

Understanding and manipulating the emergent phenomena resulting from dimensionality reduction has been a central endeavour in physics, chemistry, and materials science. Back in 1925, Ising solved in his thesis the now famous Ising model and  affirmed that there is no long-range order at finite temperatures in one dimension (1D). Mermin and Wagner~\cite{Mermin66p1133} rigorously demonstrated the absence of long-range spin or crystalline order in dimensions $d\leq 2$ using an isotropic Heisenberg model.  Dyson~\cite{Dyson69p91} later proved the existence of phase transitions in a 1D Ising model with positive and monotonically decreasing spin exchange interactions. The predicted phase transitions in 1D anisotropic Heisenberg-Ising model as well as two-dimensional (2D) XYZ model highlight  the importance of anisotropic exchange interactions~\cite{Takahashi72p2187} and slowly decaying long-range interactions~\cite{Linares99p271,Maghrebi17p023001} for the emergence of long-range order in 1D  and 2D systems. More recently, experimental and theoretical investigations  have led to the discovery of intrinsic ferromagnetism and ferroelectricity at low dimensions. Typical low-dimensional ferromagnetic materials are 2D CrI$_3$,~\cite{Huang17p270} Cr$_2$Ge$_2$Te$_6$,~\cite{Gong17p265} and Fe$_3$GeTe$_2$~\cite{Deng18p94}, while the existence of quasi-1D antiferromagnetism was reported in bulk CuCrO$_4$~\cite{Law11p014426} that possesses 1D CuO$_2$ ribbon chains. The presence of low-dimensional ferroelectricity has been demonstrated in several 2D materials such as monolayer group-IV monochalcogenides {\em MX} ($M$=Ge, Sn, $X$=S, Se)~\cite{Fei16p097601}, SnTe~\cite{Chang16p274}, $d1T$-MoTe$_2$~\cite{Yuan19p1775}, WTe$_2$~\cite{Yang18p7160}, and in 1D systems represented by SbN and BiN nanowires~\cite{Yang21p13517}.

In low-dimensional ferroelectric materials, the spontaneous polarization is often developed along a direction that is perpendicular to the direction of reduced dimension. The depolarizaiton field  resulting from the imperfect screening of the polarization bound charges at surfaces scales inversely with thickness and hence has been a main obstacle for the miniaturization of ferroelectric-based devices. Even for a perovskite ferroelectric thin film sandwiched by metallic electrodes, there could still exist a critical thickness (below which the out-of-plane polarization disappears) caused by the depolarizing electrostatic field arising from dipoles at the ferroelectric–electrode interfaces~\cite{Junquera03p506}. Among the limited number of 2D ferroelectrics, most of them possess polarization orientated in-plane. Atomically thin monolayers with out-of-plane polarization ($P_{\rm OP}$) 
have been rarely reported, and the only known ones confirmed experimentally are CuInP$_2$S$_6$~\cite{Liu1612357} and $\alpha$-In$_2$Se$_3$~\cite{Zhou17p5508}. When combined with the lateral downscaling achieved via lithographic patterning, a 2D ferroelectric with $P_{\rm OP}$ can take full advantage of the atomic thickness to realize ultrahigh-density electronic devices. For example, monolayer $\alpha$-In$_2$Se$_3$, first predicted by Ding~\etal~\cite{Ding17p14956} to be a 2D ferroelectric with first-principles density functional theory (DFT) calculations, has been demonstrated experimentally to exhibit stable $P_{\rm OP}$ at room temperatures with a thickness down to 3~nm~\cite{Xiao18p227601}. The polarization of $\alpha$-In$_2$Se$_3$ that can be manipulated by a vertical electric field allows for the fabrication of 2D ferroelectric field effect transistors (FeFETs) with a promising writing endurance of $10^5$ times~\cite{Wan18p14885, Wan19p1808606}.

The constraint on the polarization orientation imposed by the depolarization effect is carried over to 1D. There have been far fewer reports of 1D ferroelectrics than 2D and 3D ferroelectrics. The nanowire of ferroelectric BaTiO$_3$ is the most studied 1D nanostructure. Experiments showed that the smallest diameter at which the BaTiO$_3$ nanowire remains  ferroelectric at room temperatures is $\approx$ 3 nm~\cite{Spanier06p735}; first-principles calculations suggested a critical diameter of $\approx$1.2 nm,  below which the axial polarization ($P_{\rm ax}$) of the nanowire disappears~\cite{Geneste06p112906}. The nanowires of other perovskite ferroelectrics such as Pb(Zr,Ti)O$_3$ ~\cite{Hong08p064118} and KNbO$_3$~\cite{Louis10p1177} have also been synthesized. Other notable examples of 1D ferroelectricity are NbO$X_3$ ($X$=Cl, Br, and I)~\cite{Zhang21p135}, GeS, SnS~\cite{Zhang19p15040}, and WO$X_4$ ($X$ is a halogen element)~\cite{Lin19p111401}. However, for all these 1D ferroelectrics, none of them possesses polarization perpendicular to the axial direction, referred to as ``radial" polarization ($P_{\rm ra}$), due to the strong depolarization effect at the nanoscale. The mere presence of $P_{\rm ax}$ is inconvenient for lateral downscaling as a whole nanowire has to be used as the functional unit. Moreover,  the switching of  uniaxial $P_{\rm ax}$ will likely lead to high-energy charged 180$^\circ$ domain walls with head-to-head and tail-to-tail configurations. This is an unfavorable feature for long-term device reliability as those unstable domain walls may cause back-switching in the absence of external electric fields. A 1D ferroelectric with radial polarization, if available, has the potential to realize ultimately-dense ferroelectric-based electronic devices in which a few unit cells instead of a domain of unit cells serve as the functional unit.

In this work, focusing on monolayer $\alpha$-In$_2$Se$_3$ and its homologous $\alpha$-III$_2$VI$_3$ 2D sheets with $P_{\rm OP}$, we investigate the structural evolution of 1D nanoribbons with respect to the ribbon width based on DFT calculations. The 1D nanoribbon is constructed by cutting the 2D sheet with the desired edges and width. We find a spontaneous structural transition from flat nanoribbons to compact nanothreads analogous to carbon nanothreads~\cite{Chen15p14373, Fitzgibbons15p43, Xu15p5124}, as the width decreases. Most importantly, we discover a 1D ferroelectric nanothread (1DFENT) of Ga$_2$Se$_3$ (with a diameter of 6.32~\AA) that has polarization along both axial and radial directions ($P_{\rm ax}$ and $P_{\rm ra}$). The structural stability of the 1DFENT is confirmed with phonon spectrum calculations and $ab~initio$ molecular dynamic (AIMD) simulations, and its polarization switchability is validated by the nudged elastic band (NEB) method that predicts a reasonable switching barrier. Because of the strong coupling between  $P_{\rm ax}$ and $P_{\rm ra}$, both polarization variants will increase in response to a stretching axial stress, leading to an intriguing auxetic piezoelectric effect. In contrast to the conventional understanding that charge doping will destroy the long-range polar ordering, we find that both electron and hole doping can promote the magnitude of $P_{\rm ra}$ while inducing a spontaneous spin polarization, resulting in a coexistence of 1D ferroelectricity, ferromagnetism, and metallicity. Moreover, at an appropriate electron doping concentration, the 1DFENT becomes semiconducting in which the carriers are strongly localized due to the intrinsic flatness of electronic bands. The 1DFENT supporting radial polarization offers a platform for the design of ultrahigh-density ferroelectric memories and the exploration of exotic states of matter.

\section{Method}
First-principles DFT calculations are performed with \texttt{QUANTUM ESPRESSO} ~\cite{Giannozzi09p395502, Giannozzi17p465901} package using Garrity-Bennett-Rabe-Vanderbilt (GBRV) ultrasoft pseudopotentials~\cite{Garrity14p446}. The generalized gradient approximation of Perdew-Burke-Ernzerhof revised for solids (PBEsol) is chosen as the exchange-correlation functional. Vacuum layers along both $x$ and $z$ axis are set to be no less than 15 \AA ~in order to model free-standing 1D nanoribbons. The lattice constant along the axial direction ($y$-axis) and atomic positions are fully relaxed with a plane-wave cutoff of 80 Ry and a charge density cutoff of 600 Ry. A $1\times8\times1$ Monkhorst-Pack $k$-point grid is used for Brillouin zone sampling. The convergence threshold in the total energy for the ionic minimization is 10$^{-7}$ Ry and the force convergence threshold is 10$^{-6}$ Ry/Bohr. We compute the dipole moment in the unit of e\AA~per unit cell (u.c.) to gauge the magnitude of polarization.
It is noted that our previous work showed that a high-density $k$-point grid and a tight convergence threshold for electronic self-consistent calculations are needed when studying charge-doping-induced magnetic systems~\cite{Duan21p2316}. Here the convergence threshold for electronic self-consistency is set as 10$^{-11}$ Ry while the $k$-point grid remains the same as that used for structural optimization since our benchmark calculations suggested that the value of the magnetization of a charge-doped 1DFENT is well converged at a $1\times8\times1$ $k$-point grid. The optimized tetrahedron method~\cite{Kawamura14p094515} is adopted to perform $k$-space integration. The phonon spectrum of the 1DFENT of Ga$_2$Se$_3$ is calculated in the framework of density functional perturbation theory~\cite{Baroni87p1861, Gonze95p1086} with a $1 \times 12 \times 1$ $q$-point grid, and the type of acoustic sum rule is set to \texttt{crystal}. The finite-temperature structural stability is validated by performing AIMD simulations  implemented in Vienna $ab~initio$ simulation package (VASP) ~\cite{Kresse96p11169, Kresse96p15} using a $1\times4\times1$ supercell, $\Gamma$-point sampling, an energy cutoff of 350 eV, and a convergence criterion of 10$^{-5}$~eV in energy. The temperature is controlled using the Nos\'e-Hoover thermostat. The averaged structure is generated using an equilibrium trajectory of 10~ps.  We use the NEB method as well as the variable-cell NEB (VCNEB) method implemented in the \texttt{USPEX} code~\cite{Oganov06p244704,Lyakhov13p1172,Oganov11p227} to determine the minimum energy paths (MEPs) of polarization reversal. The convergence threshold in root-mean-square forces is set to 0.02 eV/\AA~ when identifying MEPs. Some key input and output files for DFT calculations are uploaded to a public repository~\cite{1DFENT_SM}.


\section{Results and Discussion}
\subsection{Structural evolution from 2D to 1D}

\begin{figure}[t]
\centering
\includegraphics[scale=0.9]{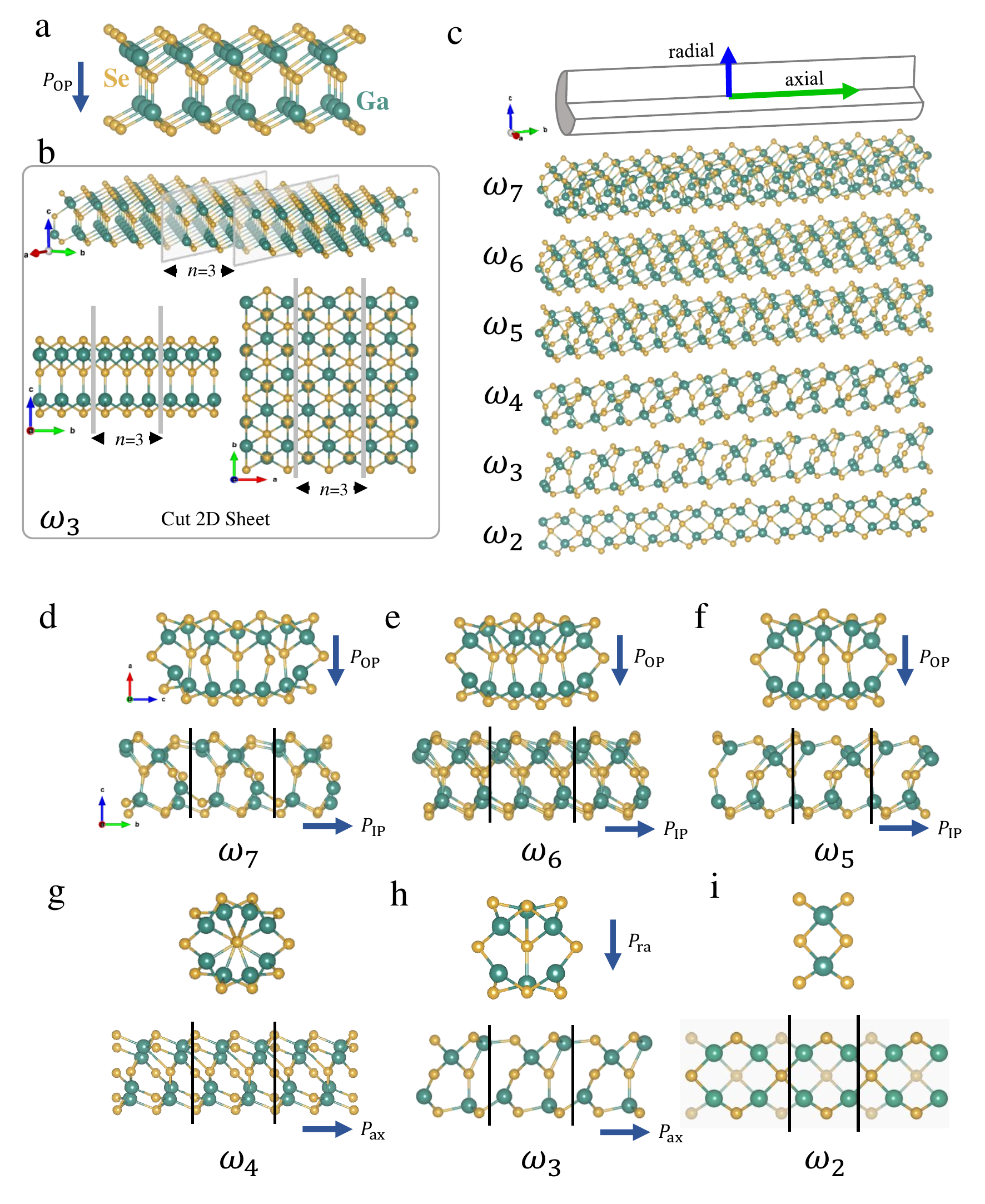}
\caption{Structural evolution from 2D to 1D in $\alpha$-Ga$_2$Se$_3$. (a) Structure of monolayer $\alpha$-Ga$_2$Se$_3$ with out-of-plane polarization ($P_{\rm OP}$). The Ga and Se atoms are represented by green and yellow balls, respectively. (b) Construction of $\omega_3$ nanoribbon by cutting the 2D sheet with armchair-terminated edges. 
The cutting planes are colored in gray. (c) Schematic illustrating the axial direction ($b$-axis) and the radial direction ($c$-axis). The bottom illustrates optimized 1D nanostructures of Ga$_2$Se$_3$ from $\omega_7$ to $\omega_2$ with axial and side views shown from (d) to (i). $\omega_3$ nanothread has both $P_{\rm ax}$ and $P_{\rm ra}$.}
\label{str}
\end{figure}

The family of $\alpha$-III$_2$VI$_3$ (III=Al, Ga, In; VI=S, Se, Te) van der Waals (vdW) materials is predicted to support $P_{\rm OP}$ in an atomic thin monolayer~\cite{Ding17p14956}. Particularly, layered $\alpha$-In$_2$Se$_3$ nanoflakes and defective Ga$_2$Se$_3$ nanosheets have been successfully synthesized in experiments~\cite{Xiao18p227601,Xue22p2105599}, both possessing 2D ferroelectricity at room temperatures.
Figure~\ref{str}a illustrates the structure of $\alpha$-III$_2$VI$_3$. The monolayer belonging to the $Pm31$ space group consists of five covalently-bonded atomic layers stacked in the sequence of VI-III-VI-III-VI, and each atomic layer has atoms arranged in a hexagonal lattice . The displacement of the central layer of VI atoms along the $z$-axis gives rise to $P_{\rm OP}$, while the switchable in-plane physical polarization is zero due to the three-fold rotational symmetry in the plane.
It is noted that according to the modern theory of polarization~\cite{KingSmith93p1651, Resta93p133, Vanderbilt93p4442}, the in-plane formal polarization of $\alpha$-III$_2$-VI$_3$ is allowed to be non-vanishing~\cite{Gibertini15p6229, Kruse22arxiv}, which could enforce gapless states at any zigzag-terminated edges~\cite{Qiao11p035431}. 


The 1D nanoribbons are constructed by cutting the $\alpha$-III$_2$VI$_3$ monolayer with different edges and widths. Similar to graphene, $\alpha$-III$_2$VI$_3$ monolayer with a hexagonal lattice can be cut either along the zigzag direction or the armchair direction. We first explored the impact of edge terminations on the stability of nanoribbons for both $\alpha$-In$_2$Se$_3$ and $\alpha$-Ga$_2$Se$_3$. Our benchmark calculations showed that the zigzag edge is unstable due to the non-vanishing formal in-plane polarization and will undergo substantial atomic reconstruction that destroys the crystalline order. In contrast, the armchair edge is stable. In view of this, we focus on 1D nanoribbons with armchair-terminated edges. Following the convention, the width $\omega_n$ is defined as the number ($n$) of armchair lines across the ribbon width (Fig.~\ref{str}b).

The structural evolution of $\alpha$-III$_2$VI$_3$ nanoribbons follows a similar trend with respect to the varying width. Here we use $\alpha$-Ga$_2$Se$_3$ as an example. We find that with reducing $n$, the 1D  Ga$_2$Se$_3$ changes gradually from a flat nanoribbon to a compact cylinder-like nanostructure, referred to as nanothread (Fig.~\ref{str}c). 
As shown in Fig.~\ref{str}d-f, 
except the notable relaxation of unsaturated edge atoms, $\omega_7$, $\omega_6$, and $\omega_5$ largely resemble the structure of monolayer $\alpha$-Ga$_2$Se$_3$ and maintain the original $P_{\rm OP}$. 
Specifically,  the $\omega_6$ nanoribbon with a width of 13.34~\AA~exhibits $P_{\rm OP}$ of 0.48 e\AA~ originated form the displacement of the central Se layer; the magnitude of in-plane polarization ($P_{\rm IP}$) is also non-zero due to the break of the in-plane three-fold rotational symmetry after the cut.
In the case of $\omega_4$, the 1D nanostructure becomes a nanothread characterized by a nearly circular base (Fig.~\ref{str}g). The reconstruction of edge atoms completely suppress $P_{\rm ra}$.
Further reducing the width to $\omega_3$ surprisingly revives $P_{\rm ra}$: the $\omega_3$ nanothread (Fig.~\ref{str}h) has both spontaneous $P_{\rm ra}$ and $P_{\rm ax}$ resulting from the displacement of the central Se atom relative to the center of the surrounding distorted Ga$_6$ octahedron (see discussions below). 
Finally, $\omega_2$ can be viewed as an ultrathin nanoribbon with neighboring four-fold coordinated Se atoms displaced in an antiparallel manner (Fig.~\ref{str}i), similar to an antiferroelectric.

\subsection{Structural stability of $\omega_3$}

Since the $\omega_3$ nanothread of III$_2$VI$_3$ is the smallest 1D system possessing both $P_{\rm ra}$ and $P_{\rm ax}$, it is natural to ask how likely could such 1D ferroelectric nanostructure be realized experimentally. A full resolution to this question with theory only is challenging. Nevertheless, we first estimate the structural stability of $\omega_3$ nanothreads by computing the phonon spectra following the general protocol of stability analysis for new materials designed computationally~\cite{Peelaers09p107}. A dynamically stable material is situated at a local minimum of the potential energy surface and will have all phonon frequencies being positive. A technical issue arises due to the 1D nature of the system: acoustic phonon modes tend have rather low frequencies (close to zero) that make the accurate determination of phonon spectrum quite difficult numerically. Using an extremely tight convergence threshold (10$^{-16}$ Ry), we obtain the phonon spectra of all $\omega_3$ nanotreads of III$_2$VI$_3$ (see Supporting Information). It is found that $\omega_3$ of Ga$_2$Se$_3$ is dynamically stable as confirmed by a phonon spectrum without imaginary vibrational frequencies (Fig.~\ref{fe}b, left panel), while $\omega_3$ of Ga$_2$S$_3$, Ga$_2$Te$_3$, and Al$_2$Te$_3$ are marginally stable. 


Additionally, we perform AIMD simulations for $\omega_3$ nanothread of Ga$_2$Se$_3$ using a $1\times4\times1$ supercell 
to investigate its structural stability at finite temperatures (Fig.~\ref{fe}b, right panel). 
The averaged structure calculated using an equilibrium trajectory of 10 ps at 200~K remains almost the same with the structure optimized with the zero-Kelvin DFT method, and the magnitude of $P_{\rm ra}$ is 0.3 e\AA/u.c., indicating a robust ferroelectricity at low temperatures. However, a higher temperature will drive structural amorphization. Further studies are needed to enhance the critical temperature for room-temperature applications. In the following, we will focus on the structural and electronic properties of $\omega_3$ nanothread of Ga$_2$Se$_3$ for its dynamical stability as well as the fact that its 2D counterpart has been synthesized~\cite{Xue22p2105599}. 

\subsection{Ferroelectric switching in $\omega_3$ of Ga$_2$Se$_3$}
\begin{figure}[t]
\centering
\includegraphics[scale=0.9]{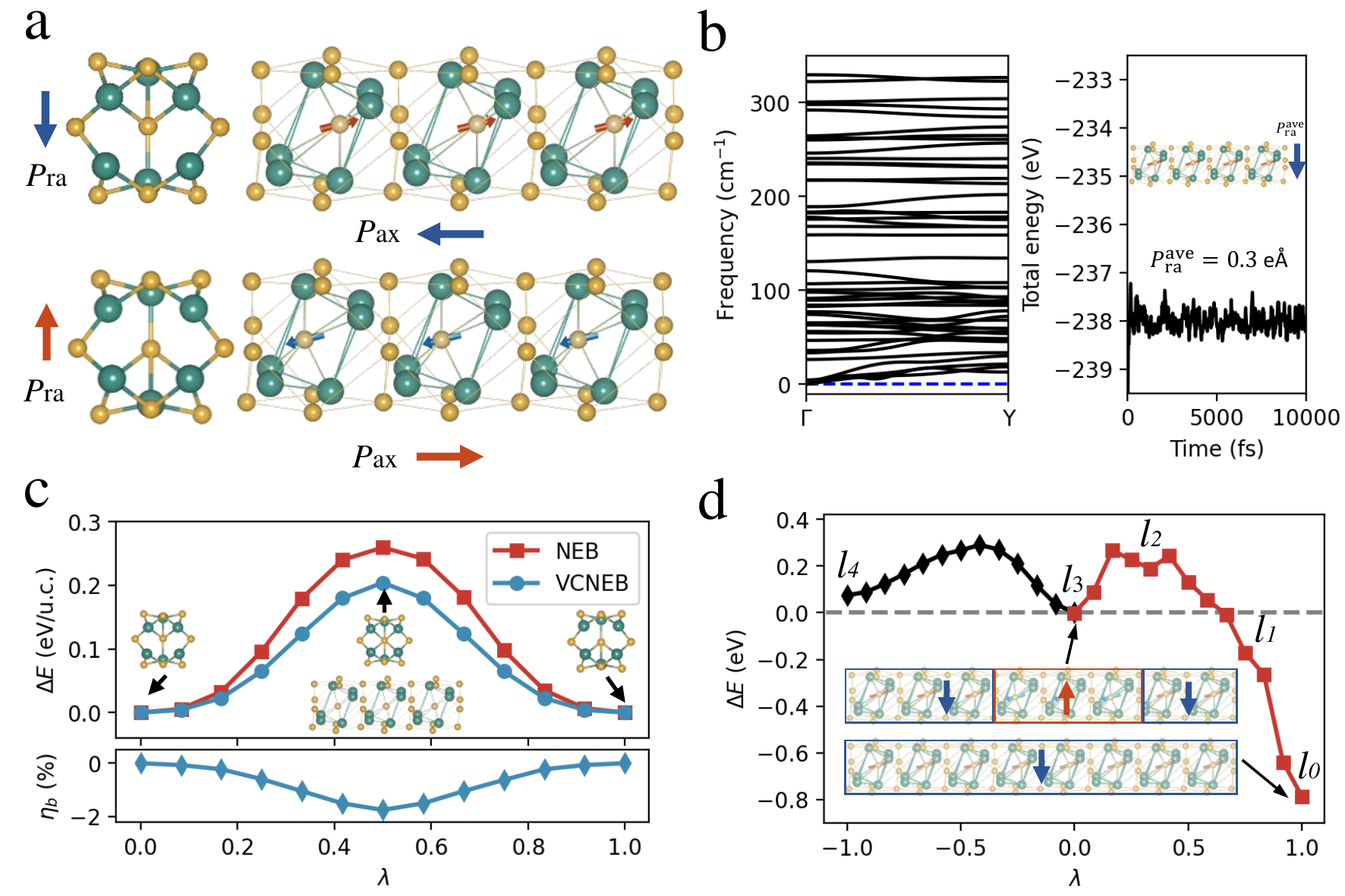}
\caption{(a) Axial and side views of two energy-degenerate polar states with opposite polarization directions in $\omega_3$ nanothread of Ga$_2$Se$_3$. The red arrow on the Se atom represents the local atomic displacement relative to the center of the Ga$_6$ cage.  
(b) Phonon spectrum (left) and energy evolution as a function of time in AIMD at 200 K (right) for the $\omega_3$ nanothread of Ga$_2$Se$_3$. The inset shows the averaged structure using an equilibrium trajectory of 10 ps. 
(c) Minimum energy paths (MEPs) of polarization reversal identified with NEB (red) and VCNEB (blue). The bottom panel shows the variation of lattice constant along the $b$ axis during the switching process, $\eta_b$, defined as $(b-b_0)/b_0$. The insets show the axial and side views of the nonpolar structure. (d) MEPs of unit-cell-by-unit-cell switching obtained with NEB. The inset shows a $l_3$ domain of upward $P_{\rm ra}$ sandwiched by domains of downward $P_{\rm ra}$.
}
\label{fe}
\end{figure}

The structural origin of the spontaneous polarization in $\omega_3$ of Ga$_2$Se$_3$ comes from the off-center displacement of Se atom relative to the center of the surrounding Ga$_6$ octahedron that is distorted along the bottom-left to top-right direction (Fig.~\ref{fe}a).   
Taking the configuration with the downward $P_{\rm ra}$ for example, the central Se atom, locating almost vertically above a bottom Ga atom, forms three shorter bonds with top-right Ga atoms and three longer bonds with bottom-left Ga atoms, thus breaking the centrosymmetry and producing both $P_{\rm ra}$ and $P_{\rm ax}$. Due to the rigidity of the Ga$_6$ octahedron, the flip of $P_{\rm ra}$ achieved via the movement of the central Se atom is inevitably accompanied by the reversal of $P_{\rm ax}$, a ``dipole locking" mechanism similar to that in monolayer $\alpha$-In$_2$Se$_3$~\cite{Xiao18p227601}. It is noted that the structural locking in $\alpha$-In$_2$Se$_3$, though being beneficial for the stabilization of $P_{\rm OP}$ against the depolarization field, makes the switching of $P_{\rm OP}$ rather difficult: the whole central Se layer must move collectively and laterally to break and form multiple In-Se covalent bonds. In comparison, the dipole locking in $\omega_3$ of Ga$_2$Se$_3$ is confined within each individual Ga$_6$ octahedron such that switching $P_{\rm ra}$ and $P_{\rm ax}$ only requires local Se movements within each unit cell, similar to the switching in perovskite ferroelectrics. This in principle will make the switching of 1DFENT easier and is a desirable feature to realize low-power ferroelectric-based devices.  The magnitudes of $P_{\rm ax}$ and $P_{\rm ra}$ computed with the Berry phase approach are 0.32~e\AA~ and 2.84~e\AA~ per unit cell, respectively. Note that $P_{\rm ra}$ of $\omega_3$ nanothread of Ga$_2$Se$_3$ is even higher than $P_{\rm OP}$ of 0.094~e\AA/u.c. in monolayer $\alpha$-In$_2$Se$_3$~\cite{Ding17p14956}.

We further evaluate the polarization switchability by mapping out the MEP of polarization reversal using the NEB method in which the axial lattice constant of each image along the switching pathway is fixed to
the ground-state value. The switching barrier obtained with NEB is 0.26 eV/u.c. (Fig.~\ref{fe}c), lower than the barrier of 0.33 eV/u.c. in tetragonal PbTiO$_3$. This suggests that the polarization can be readily reserved by applying a radial external electric field. The MEP reveals that the polarization reversal process goes through a centrosymmetric nonpolar state that has the Se atom occupying the center of the Ga$_6$ octahedron. 

Previous studies have shown that the strain relaxation effect during the ferroelectric switching can impact the barrier height~\cite{Huang22p144106}. We employ the VCNEB method to determine the MEP for the polarization reversal process that allows axial strain relaxation. The barrier estimated with VCNEB reduces to 0.2 eV/u.c. Interestingly, we observe an unusual variation in the axial strain $\eta_b$ (defined as $b/b_0-1$ with $b_0$ the ground-state lattice constant) during the switching (Fig.~\ref{fe}c, bottom panel): the nanothread will first shrink along the axial direction and then expand until it recovers to $b_0$ upon the completion of the reversal. This is distinct from the conventional ferroelectric switching process during which the
transverse dimensions perpendicular to the polar axis will first expand and then recover (contract). The contraction of the axial dimension at the initial stage of the switching in $\omega_3$ nanothread is a direct consequence of dipole locking as $P_{\rm ax}$ (and thus $\eta_b$ due to the polarization-strain coupling) must reduce together with $P_{\rm ra}$.

\subsection{Minimum information storage unit in $\omega_3$ of Ga$_2$Se$_3$}

One of the most exciting applications of ferroelectrics is for nonvolatile information storage. For a 3D bulk ferroelectric, one bit of information is often stored as a homogeneously polarized domain comprising thousands of atoms. The 1DFENT with radial polarization offers an opportunity to realize unit-cell-level data storage. To estimate the maximum theoretical data density, we first need to determine the critical length of a polar domain in $\omega_3$ of Ga$_2$Se$_3$.
A potential issue is due to the presence of axial polarization that the 180$^\circ$ domain walls separating domains with opposite $P_{\rm ra}$ will acquire positive (negative) bound charges due to $P_{\rm ax}$ directed toward (from) the wall. We calculate energy barriers for unit-cell-by-unit-cell switching with NEB and identify a critical length of three unit cells ($l_3$). As shown in Fig.~\ref{fe}d, the switching of one unit cell that changes a $l_4$ domain to a $l_3$ domain or vice versa needs to overcome a barrier of 0.28 eV, large enough to prevent spontaneous back-switching. However, smaller polar domains such as $l_2$ and $l_1$ are no longer stable, and they will spontaneously adopt the polar state of neighboring larger domains.

We now give a rough estimation to the maximum theoretical data density enabled by 1DFENTs with radial polarization. 
We obtain the density of packed 1DFENTs by optimizing a cell containing four aligned $\omega_3$ nanothreads. The equilibrium distance between neighboring nanothreads is $\approx$3.5~\AA. Assuming the $l_3$ domain being the minimum storage unit, the theoretical data density for an atomically thin layer of 1DFENT array is $\approx$600 Gigabit/mm$^2$ (see calculations in Supporting Information), 60 times higher than current NAND areal density ($\approx$10 Gigabit/mm$^2$)~\cite{Park21p20528852}. Furthermore, by packing 1DFENTs into 3D arrays, the volumetric density could approach 600 Petabit/mm$^3$. Noted that these values are upper bounds since contacting electrodes are not considered.

\subsection{Piezoelectric properties of $\omega_3$ of Ga$_2$Se$_3$}

\begin{figure}[t]
\centering
\includegraphics[scale=0.9]{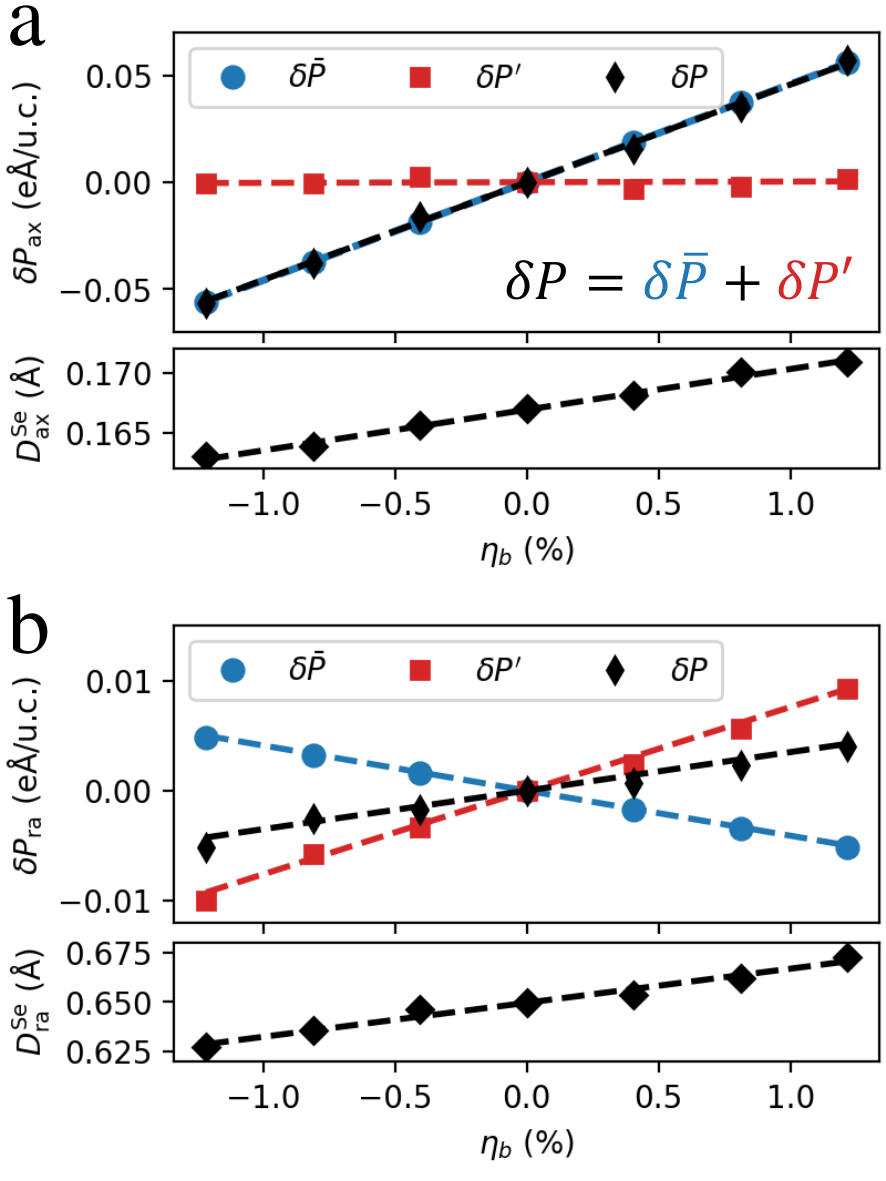}
\caption{Piezoelectricity of $\omega_3$ Ga$_2$Se$_3$. Variation of (a) axial polarization ($\delta P_{\rm ax}$) and (b) radial polarization ($\delta P_{\rm ra}$) and local displacements of Se atoms ($D^{\rm Se}$) as a function of axial strain ($\eta_b$). The change in polarization is decomposed into clamped-ion contribution ($\delta \bar{P}$) and internal-strain contribution ($\delta P'$). 
}
\label{piezo}
\end{figure}

The dipole locking in the $\omega_3$ nonothread hints at unusual piezoelectricity. We compute the piezoelectric stress coefficients $e_{32}$ and $e_{22}$ from the slopes of polarization versus axial strain curves (Fig.~\ref{piezo}) and find that both coefficients are of the positive sign, an indication of auxetic piezoelectric effect (APE) that stretching the nanothread will enhance both $P_{\rm ra}$ and $P_{\rm ax}$. This is in sharp contrast with traditional piezoelectric materials in which the longitudinal and the transverse piezoelectric coefficients are of opposite signs. The auxetic piezoelectric effect can be viewed as a piezoelectric
analogy of auxetic materials characterized by a negative Poisson's ratio. 


To understand the origin of APE, we decompose the axial strain-induced polarization change ($\delta P$) into the clamped-ion contribution ($\delta\bar{P}$) computed with the internal atomic coordinates fixed at their zero-strain values and the internal-strain contribution ($\delta P'$) arising from ion relaxations, $\delta P= \delta \bar{P} + \delta P'$. Similar decomposition can be made to the total piezoelectric coefficient, $e_{ij}= \bar{e}_{ij} + e'_{ij}$, using $e_{ij}=\partial{P_{i}}/\eta_j$ ~\cite{Szabo98p4321, Liu17p207601}.
We find that for the axial polarization, the clamped-ion term makes dominate contribution ($\delta P_{\rm ax}\approx\delta \bar{P}_{\rm ax}$) while the internal-strain contribution is close to zero, indicating that the electronic relaxation is responsible for the axial piezoelectric response $e_{22}$. Interestingly, for the radial polarization, $\delta\bar{P}_{\rm ra}$ is negative due to a tensile axial strain ($\eta_b>0$), leading to a negative clamped-ion $\bar{e}_{32}$. The total positive response of $e_{32}>0$ is resulting from the positive internal-strain contribution of $e_{32}'>0$. Structurally, the APE is a manifestation of dipole locking:  the tensile axial strain will promote the Se displacement along the axis ($D_{\rm ax}^{\rm Se}$) as well as the ``locked" radial displacement ($D_{\rm ax}^{\rm Se}$), thus enhancing both $P_{\rm ra}$ and $P_{\rm ax}$.

\subsection{Electronic properties of $\omega_3$ of Ga$_2$Se$_3$}
\begin{figure}[t]
\centering
\includegraphics[scale=0.9]{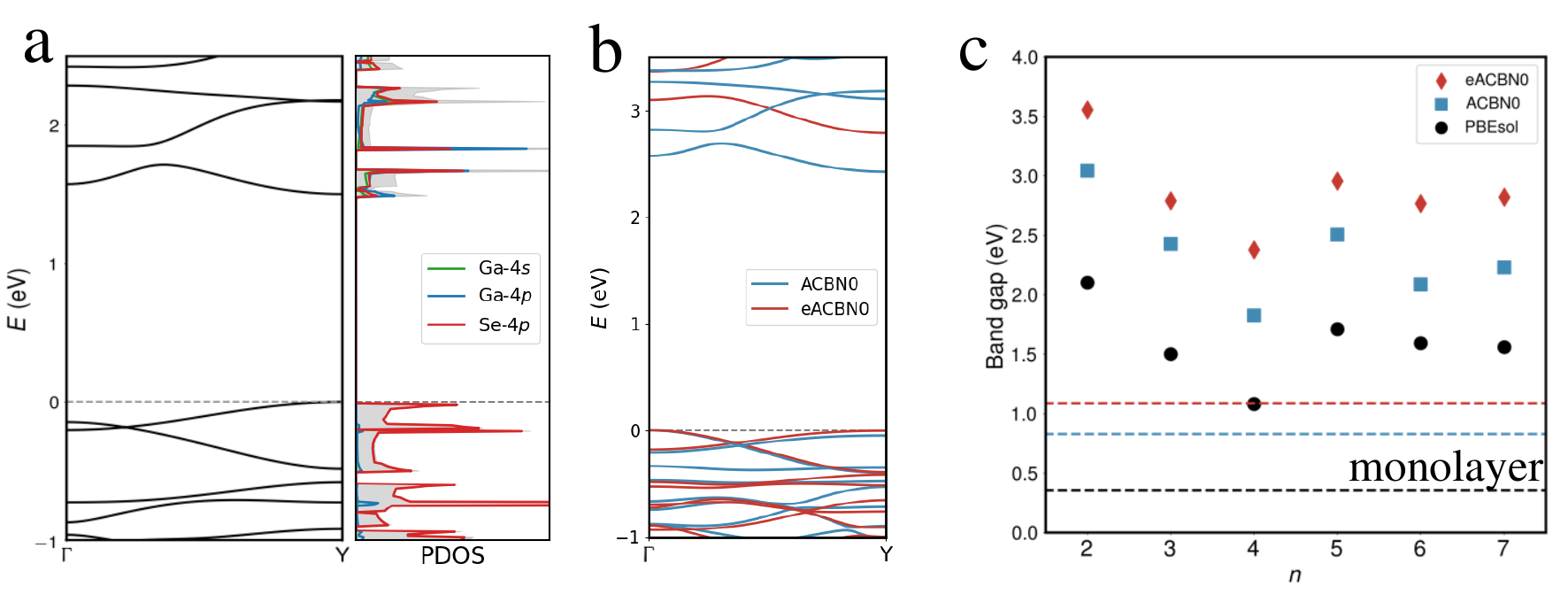}
 \caption{Electronic structure of $\omega_3$ Ga$_2$Se$_3$. (a) Band structure and projected density of states computed with PBEsol. (b) Band structures computed with ACBN0 and eACBN0. (c) Band gap as a function of width $n$ of 1D nanostructure of Ga$_2$Se$_3$.}
  \label{ele}
 \end{figure}

The PBEsol band structure of $\omega_3$ of Ga$_2$Se$_3$ is presented in Fig.~\ref{ele}a, revealing a direct band gap of 1.5~eV at Y. The projected density of states (PDOS) show that valence band maximum (VBM) takes almost exclusively a Se-4$p$ character while the conduction band minimum (CBM) consists of Se-4$p$ and Ga-4$p$ characters. Most bands have small bandwidths. For example, the bandwidth of the highest valence band is only 0.18 eV, indicating the contributing Se-4$p$ orbitals are mostly localized with weak inter-orbital hybridizations. 

Since the semi-local density functionals like PBEsol often underestimate the band gap, we further compute the band structure using the newly developed pseudohybrid Hubbard density functional, Agapito–Cuetarolo–Buongiorno Nardelli (ACBN0)~\cite{Agapito13p165127} and the extended version (eACBN0)~\cite{Lee20p043410, Tancogne-Dejean20p155117}. The ACBN0 function is essentially a DFT+$U$ method with Hubbard parameter $U$ computed self-consistently. The eACBN0, a DFT+$U$+$V$ method, takes into account the intersite Coulomb interaction $V$ between neighboring Hubbard sites. Previous studies have demonstrated that for a wide range of materials ACBN0 and eACBN0 have improved accuracy over PBE and are on par with advanced methods such as the Heyd-Scuseria-Ernzerhof (HSE) hybrid density functional and
$GW$ approximations but at a lower cost~\cite{Huang20p165157, Ke21p3387, Yang22p195159, Duan21p2316, Lee20p043410}.  Particularly for low-dimensional materials, the reliability of HSE that assumes a fixed dielectric screening is not justified~\cite{Jain11p216806}. The inclusion of self-consistent Hubbard parameters helps to capture the rapid variation in Coulomb screening at low dimensions. The ACBN0 and eACBN0 band structures of $\omega_3$ nanothread of Ga$_2$Se$_3$ are presented in Fig.~\ref{ele}b, and the predicted band gaps are 2.47 and 2.79 eV, respectively. We further calculate band gaps for all 1D nanostructures of Ga$_2$Se$_3$ discussed in Figure 1. The eACBN0 method consistently predicts a larger band gap than ACBN0, followed by PBEsol, and all nanostructures  have larger band gaps than monolayer $\alpha$-Ga$_2$Se$_3$ (0.35 eV for PBEsol, 0.83 eV for ACBN0, and 1.09 eV for eACBN0, respectively).



\subsection{Coexistence of 1D ferroelectricity and ferromagnetism in $\omega_3$}

The narrow electronic bands in $w_3$ of Ga$_2$Se$_3$ lead to high values of density of states. According to the Stoner criterion, it is possible to induce Stoner-type magnetism by adjusting the Fermi level ($E_F$) to the energy level associated with a high density of states via appropriate charge doping ~\cite{Cao15p236602, Seixas16p206803}. If the charge doping does not destroy the long-range polar ordering, the doped 1D system could host both ferroelectricity and ferromagnetism.

\begin{figure}[t]
\centering
\includegraphics[scale=0.8]{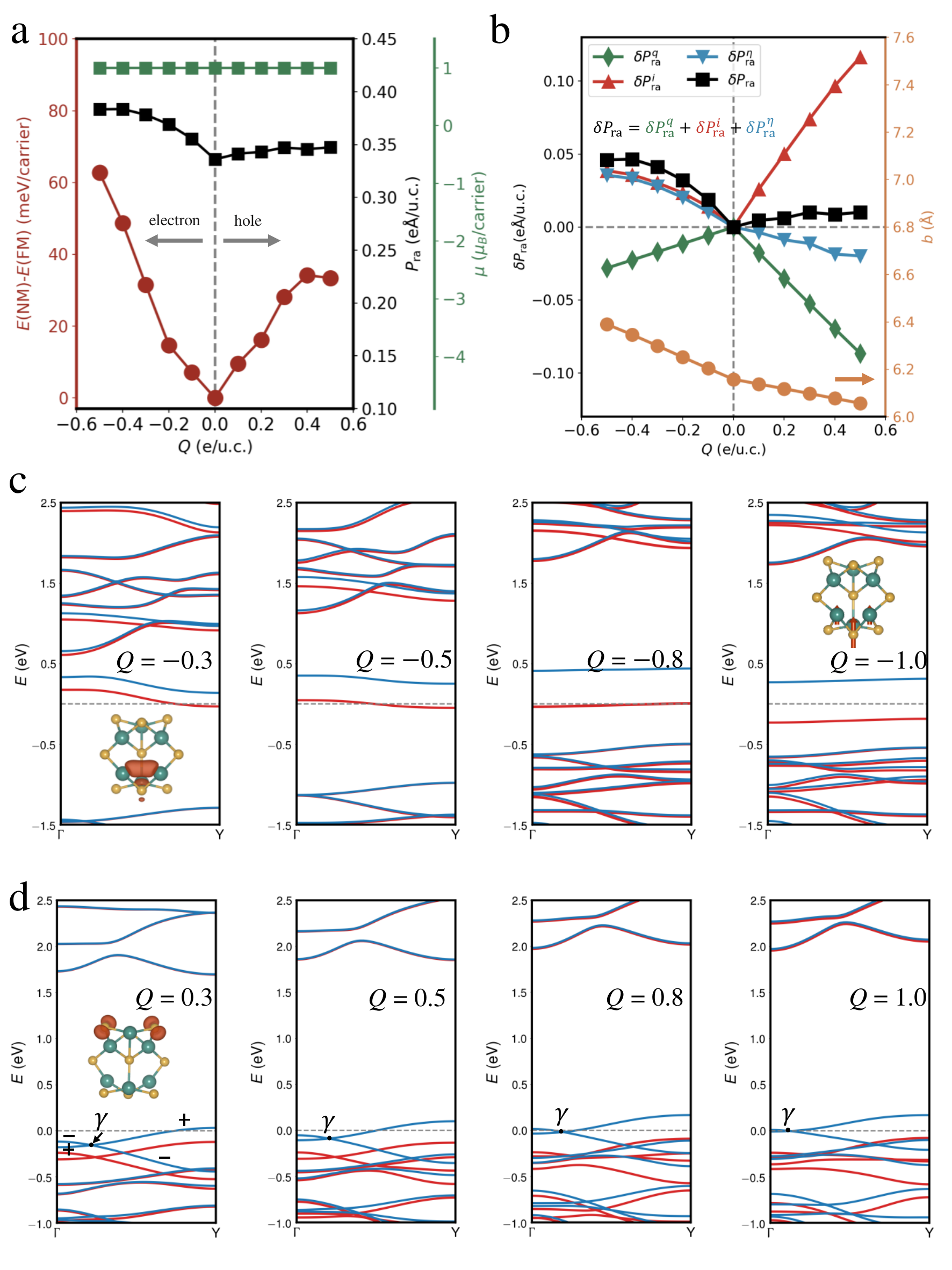}
 \caption{Ferromagnetism of charge-doped $\omega_3$ Ga$_2$Se$_3$. (a) Spin polarization energy $E$(NM)-$E$(FM), radial polarization, and magnetic moment ($\mu$ in Bohr magneton $\mu_B$ per carrier) as a function of doping concentration ($Q$). (b) Doping-induced change in the radial polarization ($\delta P_{\rm ra}$) and lattice constant $b$ as a function of $Q$. $\delta P_{\rm ra}$ is decomposed into charge-carrier contribution ($\delta P_{\rm{ra}}^{q}$), ion relaxation contribution ($\delta P_{\rm{ra}}^{i}$), and strain contribution ($\delta P_{\rm{ra}}^{\eta}$). Spin-polarized band structures for (c) electron and (d) hole doping. The insets show the spin-polarized charge density isosurfaces. }
  \label{mag}
 \end{figure}

We examine the ferroelectric and magnetic properties of $\omega_3$ as a function of charge-carrier concentration ($Q$ in the unit of electron/hole per unit cell) at the PBEsol level. 
As shown in Fig.~\ref{mag}a, our results reveal the emergence of a spontaneous time-reversal symmetry breaking in both electron-doped and hole-doped $\omega_3$ over a wide range of doping concentration from $Q=-0.5$ to $0.5$~e/u.c.
The doping-induced ferromagnetism is further confirmed by calculating the spin polarization energy defined as the energy difference between the nonmagnetic (NM) state and the ferromagnetic (FM) state, E(NM)-E(FM). We find that the spin polarization energy remains positive at all studied doping concentrations and increases with increasing magnitude of $Q$, indicating a strong magnetic instability at the nonmagnetic state that drives the spontaneous break of time-reversal symmetry.
The value of averaged magnetic moment per carrier ($\mu$) has a constant value of 1 independent of $Q$. This shows that all doped carriers are spin polarized. 

In bulk ferroelectrics, the metallicity and ferroelectricity are often mutually exclusive since charge carriers will screen long-range Coulomb interactions that drive the structural distortion~\cite{Wang12p247601}.  An important question here is whether the long-range polar ordering in $w_3$ nanothread could survive the charge doping. Unexpectedly, as shown in Fig.~\ref{mag}b, both electron and hole doping will enhance the magnitude of $P_{\rm ra}$. Following a similar protocol in ref.[55,56], we perform three calculations in order to identify and separate contributions to the charge-carrier-induced change in $P_{\rm ra}$ relative to the undoped value.
Starting with the undoped nanothread, we first compute the change in $P_{\rm ra}$ after introducing charge carriers with atomic positions and lattice constants fixed; we refer to this polarization change as the charge-carrier contribution, $\delta P_{\rm ra}^q$. Then we relax the atomic positions with fixed lattice constant $b$; the further induced polarization change is called ion relaxation contribution, $\delta P_{\rm ra}^i$. Finally, the lattice constant $b$ and the atomic positions are fully relaxed; we name this third component as strain contribution, $\delta P_{\rm ra}^\eta$. Therefore, the total change in polarization is decomposed as $\delta P_{\rm ra} = \delta P_{\rm ra}^q + \delta P_{\rm ra}^i + \delta P_{\rm ra}^\eta$ .
The $Q$-dependence of $\delta P_{\rm ra}^q$ follows the conventional understanding that charge carriers will suppress the polarization: both hole and electron doping give negative values of $\delta P_{\rm ra}^q$. In contrast, the ion-relaxation contribution promotes $P_{\rm ra}$ as all values of $\delta P_{\rm ra}^i$ are positive. The strain contribution follows the APE discussed above: the electron doping increases $b$ thus leading to positive $\delta P_{\rm ra}^\eta$ whereas the hole doping reduces $b$ and causes negative $\delta P_{\rm ra}^\eta$. Overall, the polarization enhancement along the radial direction is because of both ion relaxation and strain contributions in the electron doped region but solely due to the ion relaxation in the hole doped region.

\subsection{Charge doping induced electronic transition}

The evolution of the spin-polarized band structures as a function of the concentration of doped electrons reveals a doping-driven band flattening: the bandwidth ($t$) of the spin-up and spin-down bands near the $E_F$ reduces from 160~meV at $Q=-0.3$ to 45~meV at $Q=-1.0$ (Fig.~\ref{mag}c). By the meanwhile, the energy difference between the two bands, roughly viewed as the strength of on-site Coulomb repulsion ($U$),  increases from 2.95 eV to 3.14 eV (for Se-4$p$ orbitals).
The electron-doped $\omega_3$ of Ga$_2$Se$_3$ is charge-carrier-mediated ferromagnetic metal before reaching the critical concentration of $Q=-1$ at which the system becomes a fully gaped insulator. Different from the band insulator at $Q=0$, the $Q=-1$ system can be viewed as a 1D two-level system at half filling with $U/t\gg1$ and is an interaction-induced insulator in which each unit cell hosts a localized doped electron, resembling the Hund's rule for orbital filling. Because of the small $t$, the exchange coupling between neighboring spins is expected to be weak. Consequently,the antiferromagetic state will have energy comparable with the ferromagnetic state at 0~K. This is confirmed with DFT calculations that show the antiferromagnetic state is slightly higher in energy by 0.7~meV per unit cell than the ferromagnetic state. At an elevated temperature, the $\omega_3$ nanothread at $Q=-1$ is most likely a paramagnetic state due to the weak exchange interaction.

In comparison, the hole doping impacts the band dispersion to a lesser degree, but enhances the exchange splitting between the spin-up and spin-down bands. Moreover, we find that the hole-doped $\omega_3$ is a ferromagnetic metal protected by the mirror reflection symmetry with respect to the $xy$ plane ($\mathcal{M}_z$).  The highest two valence bands exhibit  opposite $\mathcal{M}_z$
 parities as marked by +1 and $-1$ in Fig.~\ref{mag}d. The crossing point (labeled as $\gamma$) between the two bands belonging to different irreducible representations is symmetry protected and will not hybridize to open a gap regardless the hole doping concentration. Consequently, the hole-doped $\omega_3$ is a symmetry-protected ferromagnetic metal. 

\section{Conclusions}
In summary, we predict a 1D ferroelectric nanothread of ultrasmall diameter that exhibits both axial and radial polarization based on first-principles density functional theory calculations. We show that there is a spontaneous structural evolution from nanoribbon to nanothread with reducing nanoribbon width. The switchable radial polarization afforded by the $\omega_3$ nanothread of Ga$_2$Se$_3$ has the potential to realize ultimately-dense ferroelectric-based electronic devices in which a 1D domain of just three unit cells could serve as the functional unit.
The dipole locking feature gives rise to an intriguing 1D auxetic piezoelectric effect that the axial stretching promotes both radial and axial polarization. Moreover, the 1D ferroelectricity is robust against both electron and hole doping, enabling the coexistence of ferroelectricity, ferromagnetism, and metallicity. The intrinsic flatness of electronic bands in 1D nanothread offers a platform to explore exotic states of matter such as Hund's insulator-like semiconductor. 
Our DFT calculations offer theoretical evidences supporting the structural stability of proposed 1D ferroelectric nanothreads.
We hope the proof-of-concept reported in this work will motive experimental studies toward the synthesis of this new family of low-dimensional ferroelectrics. 


\section{Supporting Information}
Phonon spectra of $\omega_3$ nanothreads of III$_2$VI$_3$ (Sect.~I), Packed $\omega_3$ nanothreads (Sect.~II). This material is available free of charge via the Internet at \url{https://pubs.acs.org}.
\begin{acknowledgement}

J.H., C.K., and S.L. acknowledge the supports from Westlake Education Foundation, and Westlake Multidisciplinary Research Initiative Center, and National Natural Science Foundation of China (52002335).

\end{acknowledgement}

\bibliography{SL}
 
\clearpage
\newpage

\clearpage
\newpage

\clearpage
\newpage

 \clearpage
 \newpage
 
 \clearpage
 \newpage
 
 
\end{document}